\documentclass[aps,prc,floatfix,amsmath]{revtex4}
\usepackage{dcolumn}
\usepackage{mathrsfs}
\usepackage{supertabular}
\usepackage{color,graphicx}
\usepackage[bookmarksopen]{hyperref}
\begin{document}
\title{Energy and momentum relaxation of heavy fermion in dense and warm plasma}
\author{Sreemoyee Sarkar}
\email{sreemoyee.sarkar@saha.ac.in}
\author{Abhee K. Dutt-Mazumder}
\email{abhee.dm@saha.ac.in}
\affiliation{High Energy Physics Division, Saha Institute of Nuclear Physics,
1/AF Bidhannagar, Kolkata-700 064, INDIA}

\medskip

\begin{abstract}
We determine the drag and the momentum diffusion coefficients of heavy 
fermion in dense plasma. It is seen that in degenerate matter drag coefficient at the leading order  
mediated by transverse photon is proportional to $(E-\mu)^2$ while for the 
longitudinal exchange this goes as $(E-\mu)^3$. We also calculate the longitudinal
diffusion coefficient to obtain the Einstein relation in a relativistic degenerate 
plasma. 
Finally, finite temperature corrections are included both for the drag and the diffusion coefficients. 
\end{abstract}

\maketitle

%%%%%%%%%%%%%%%%%%%%%%%%%%%%%%%%%%%%%%%%%%%%%%%%%%%%%%%%%%%%%%%%%%%%%%%%%
\section{Introduction}
%%%%%%%%%%%%%%%%%%%%%%%%%%%%%%%%%%%%%%%%%%%%%%%%%%%%%%%%%%%%%%%%%%%%%%%%%

Recent years have witnessed significant progress in understanding the 
properties of hot and/or dense relativistic plasma 
\cite{BELLAC, KAPUSTA}. Such studies draw their motivations both from the theory and the experiments. In particular, the possibility
of creating high temperature quark gluon plasma (QGP) by colliding heavy ions
in the laboratory mimicking the conditions of microsecond old universe has
been a matter of intense research activities in the past decades. 
Further impetus to these studies comes from astrophysics where it is important 
to know the properties of such plasma at high density, which, for example,
 might exist in the core of neutron stars or in white dwarfs.  

One of the interesting quantities which has assumed special interest recently
is the study of partonic energy loss in relativistic plasma. Several 
calculations \cite{BT1, BT2, ANDRE1, ANDRE2, AKDM} have been performed over the last 
decades to estimate such energy loss in a plasma. Similarly, there exists 
several studies in which momentum diffusion coefficient of heavy fermion 
has been estimated \cite{BENJAMIN, GDM, ABHEEP, BERAUDO}. These two quantities
are of utmost importance to understand the equilibration of fermions
in a plasma. So far, these calculations were largely confined to the case 
of hot plasma with zero chemical potential
due to their relevance to the experiments performed at the Relativistic Heavy
Ion Collider (RHIC) or the ones to be performed  at the 
Large Hadron Collider (LHC).

There still exists another domain of Quantum Chromodynamic (QCD) phase diagram
where the chemical potential ($\mu$) might be higher compared to the temperature ($T$).
This is the region of interest of the upcoming experiments on compressed 
baryonic matter (CBM) to be performed at FAIR/GSI \cite{TOLOS, SENGER1, SENGER2}. Partially
motivated by these proposed experiments and partly by another theoretical work 
on fermion damping rate \cite{CRISTINA}, we calculate here the drag ($\eta$) and the
longitudinal momentum diffusion coefficient (${\cal B}$)  of a  heavy fermion 
in Quantum Electrodynamic (QED) plasma. It is known that
the former and the latter are related to the energy loss and the momentum relaxation
of the fermion in a plasma. Moreover, in equilibrating plasma, these two
quantities {\em viz.} $\eta$ and ${\cal{ B}}$ are related to each other
{\em via} Einstein relation (ER) which at finite temperature reads as
${\cal {B}}=2ET\eta$. As indicated above, such calculations, for dense ($T=0$) and/or warm ($T\ll \mu$) plasma
are rather limited. In fact, we are aware of only one calculation of
energy loss where the effects of finite chemical potential has been
considered, although the temperature considered there is still high \cite{VIJA}. We on the contrary first consider the extreme case of zero temperature
and then incorporate finite temperature corrections to our result
both for the drag (energy loss) and the diffusion coefficient in the limit $\mu\gg T$. We also
determine the relationship between $\eta$ and ${\cal B}$
\textit{i.e.} ER at zero temperature, which shows some interesting behavior due
to finite density plasma effect.

Before we proceed further, it would be  worthwhile to draw our
attention to \cite{CRISTINA}. This is an interesting work in many
ways. First, it is known that the fermion damping rate ($\gamma$) 
in hot plasma is 
plagued with divergences which cannot be removed by the ordinary 
screening effect \cite{BP}. This is because, the magnetic interaction is
screened only dynamically \cite{WELDON} and the problem remains for the static 
photons (or gluons in QCD). Therefore, to obtain
finite result, a suitable resummation has to be performed. This was first done 
in \cite{BIANCU1, BIANCU2}.
Ref.\cite{CRISTINA} shows that at zero temperature due to Pauli Blocking,
finite result can be obtained without performing further resummation. This
is consistent with the conclusion drawn in \cite{VO}. Secondly, in the relativistic plasma $\gamma$ is dominated by the magnetic exchange and is proportional to
$(E-\mu)$, while the electric photon exchange gives a contribution
proportional to $(E-\mu)^2$. Here, it is important
to note that the dynamical screening in the transverse sector enhances the damping rate compared
to its longitudinal counterpart. 
It might be recalled also that for non-relativistic Coulomb
plasma the damping rate goes as $(E-\mu)^2$ \cite{QF}. 
Thus, it would be interesting
to see how do the drag and the diffusion coefficient
depend on $(E -\mu)$ in degenerate plasma. 

It is known that at finite temperature
the calculation for the energy 
loss and diffusion coefficients are plagued with infrared (IR) divergences
\cite{BP}. To deal with this problem, in hot plasma one separates the
integration into two domains: one involving the exchange of hard photons 
(or gluons) {\em i.e.} the momentum transfer ($q$) $\sim T$ and the other 
involving soft photons (or gluons) when $q\sim eT$ ($e\ll 1$). 
In case of the former, one uses bare propagator and introduces an arbitrary 
cut off ($q^*$) \cite{BY} parameter to regularize the integration. For the latter, 
on the other hand one uses the 
Hard Thermal Loop (HTL) corrected propagator. 
These two parts, upon addition, yield
results independent of this intermediate scale.
In case of degenerate plasma, one also encounters similar infrared
divergences and following ref.\cite{CRISTINA} one can proceed along
the same way as finite temperature (using Hard Dense Loop (HDL) corrected propagator) and show that both for the drag
and diffusion coefficient the final result becomes independent of
the intermediate cut-off parameter. This however, as we shall see, is
not required in case of dense plasma. Here, the dominant or the leading
order contribution comes entirely from the soft sector and the hard
photon exchange contributes only to the higher order. It might be mentioned here that
although we calculate these quantities for QED, with appropriate 
color factors the results can easily be extended to the case of QCD with
the addition of one more diagram involving triple 
gluon vertex \cite{BT2}. Furthermore, it might be noted that the quark energy
loss calculations in general should also include Bremsstrahlung radiation
of the gluons. However, in the present context we are concerned with only
the two body scatterings and therefore restrict ourselves to the collisional
energy loss alone. 

Furthermore, expressions derived for the degenerate plasma, wherever possible, 
have been directly compared with
their finite temperature counterparts (with zero chemical potential). This 
brings the similarities and the differences of these two extreme
scenarios into clearer relief. 

The plan of the paper is as follows. First in section II we calculate
drag and diffusion coefficients in degenerate plasma and discuss about Einstein relation. In section III,
the finite temperature corrections have been incorporated both for $\eta$ and ${\cal B}$. The results
are then summarized in section IV. An appendix has also been added to understand
the origin of difference in ER in cold medium than from that
of hot plasma.

%%%%%%%%%%%%%%%%%%%%%%%%%%%%%%%%%%%%%%%%%%%%%%%%%%%%%%%%%%%%%%%%%%%%%%
\section{Heavy fermion at zero temperature}
\subsection{Drag coefficient}
%%%%%%%%%%%%%%%%%%%%%%%%%%%%%%%%%%%%%%%%%%%%%%%%%%%%%%%%%%%%%%%%%%%%%%
\begin{figure}[htb]
\begin{center}
\resizebox{8.5cm}{4.75cm}{\includegraphics{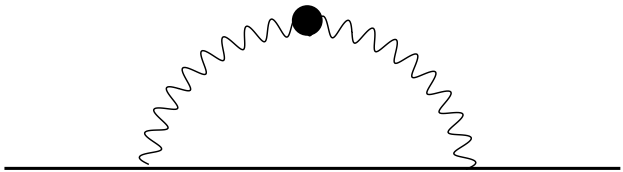}}
\caption{Fermion self-energy with resummed photon propagator.
\label{fig1}}
\end{center}
\end{figure}

In this section we first calculate the drag coefficient of a heavy fermion in 
a degenerate QED plasma. For this we consider scattering of a heavy 
fermion having energy $(E)$ (which we assume to be hard), with the constituents of the plasma 
{\em viz.} the electrons. Incidentally, this drag co-efficient ($\eta$) is
related to the to the energy loss by the following equation: 
\begin{eqnarray}
\eta&=& {1\over E v_i}\Big(-{dE\over dx}\Big),
\label{eqno(32)}
\end{eqnarray}
where, $\textbf{v}_i={\textbf{p} \over E}$ is the velocity of the incident 
fermion, $(dE/dx)$ is the 
energy loss and ${\bf p}$ is the three momentum of the incident fermion. Thus,
the calculation of the drag coefficient boils down to the calculation of
collisional energy loss in a plasma \cite{BT1, BT2, ANDRE1, ANDRE2, AKDM}. Now the
energy loss $(dE/dx)$ can be 
obtained by averaging over the
 interaction rate times the energy transfer per scattering $\omega$ and dividing by the velocity of the incoming
 particle \cite{BT1},
\begin{equation}
 {dE\over dx}={1\over v_i}\int d\Gamma \omega.
\label{eqno(1)}
\end{equation}
 
This expression is quite general and valid for both the finite temperature
and/or density where only the phase space will be different
due to the modifications of the distribution functions depending upon
the values of $\mu$ and $T$.

The scattering rate, which is essential for the calculation of $\eta$ as evident from Eqs.(\ref{eqno(1)}) and (\ref{eqno(32)}) is
related to the imaginary part of the fermion self energy ($\Sigma$) by
the following equation \cite{WELDON2}:

\begin{equation}
 \Gamma(E)=-{1\over 2 E}  {\rm Tr}\,\left[{\rm Im}\,\Sigma(p_0+i\eta\ ,{\bf
p})({P\llap{/\kern1pt}}+m)\right] \Big|_{p_0=E}  \ .
\label{eqno(2)}
\end{equation}
In the last equation, $m$ is the mass of the incoming heavy fermion. 
The full fermion self-energy represented in Fig.(\ref{fig1}) can be written explicitly 
as: 
\begin{equation}
\Sigma(P)= e^2T\sum_s\int{{\rm d}^3q\over (2\pi)^3}\gamma_\mu \,
S_f(i(\omega_n-\omega_s),{\bf p-q})\gamma_\nu \,
\Delta_{\mu\nu}(i\omega_s,{\bf
q}) \ ,
\label{eqno(3)}
\end{equation}

where, $p_0=i\omega_n+\mu$, $q_0=i\omega_s$. $\omega_n=\pi(2 n+1)T$ and 
$\omega_s=2\pi s T$ are the Matsubara frequencies for fermion and boson 
respectively with integers $n$ and $s$. After performing the sum over Matsubara frequency in Eq.(\ref{eqno(3)}), $i\omega_n+\mu$ is analytically continued 
to the Minkowski space $i\omega_n+\mu\rightarrow p_0+i\eta$, with 
$\eta\rightarrow0$. The blob in Fig.(\ref{fig1}) here represent HTL/HDL corrected photon propagator which is in the 
Coulomb gauge is given by \cite{BELLAC},

\begin{eqnarray}
\Delta_{\mu\nu}(Q)=\delta_{\mu 0}\delta_{\nu 0} \, \Delta_l(Q)
+{P}^t_{\mu \nu}\Delta_t(Q) \ ,
\label{eqno(6)}
\end{eqnarray}

with, ${P}^t _{i j} = (\delta_{ij}-\hat q_i\hat q_j), {\hat q}^i = {\bf q}^i/|{\bf q}|$, 
  ${P}^t_{i0} = {P}^t_{0i} =
{P}^t_{00}=0$ and $\Delta_{l}$, $\Delta_{t}$ are given by \cite{BELLAC},
\begin{eqnarray}
\Delta_{l} (q_0, q)   & = & {-1\over q^2+\Pi _{l}},  \\
 \Delta_{t}  (q_0, q) &  =  & {-1\over q_0^2-q^2-\Pi _{t}} .
\label{eqno(7)}
\end{eqnarray}
For subsequent calculations it is convenient here to introduce the spectral functions $\rho_{l,t}$ \cite{BELLAC}:
\begin{eqnarray}
\rho_{l,t} (q_0,\textbf{q}) &=& 2{\rm Im} \Delta_{l,t} (q_0+i\eta, \textbf{q}) .
\label{eqno(7a)}
\end{eqnarray}
At the leading order these are derived from the one-loop photon self-energy where the loop momenta 
are assumed to be hard in comparison to the photon momentum \cite{MANUEL2, MANUEL3}. In the literature the formalism is known as the HTL/HDL 
approximation as discussed in \cite{BELLAC},
\begin{eqnarray}
 \rho_{l}(q_0,q) & =&
 \frac{2 \pi m_{D}^2\,  x \,\Theta(1-x^2)}{2
\left[ q^2 +m_{D}^2  \left( 1 - \frac{x}{2} 
\ln {\Big| \frac{x+1}{x-1} \Big|} \right) \right]^2 + \frac{m_{D}^4  \pi^2 x^2 }{2}
} \ , \nonumber\\
\rho_{t}(q_0,q) & =&
 \frac{2 \pi m_{D}^2\, v_f^2\, x \,(1-x^2)\Theta(1-x^2)}{
\left[2 q^2(x^2 v_f^2 -1) -m_{D}^2 x^2 v_f^2 \left( 1 + \frac{(1-x^2)}{2x} 
\ln {\Big| \frac{x+1}{x-1} \Big|} \right) \right]^2 + \frac{m_{D}^4 v_f^4 \pi^2 x^2 (1- x^2)^2}{4}}, \ 
\label{eqno(9)}
\end{eqnarray}
where, $v_f$ is the Fermi velocity and $x={q_{0}\over qv_f}$. For a ultrarelativistic plasma ($v_f\rightarrow 1$) the Debye
mass is $m_D^2= {e^2 \over \pi^2}\Big(\mu^2+{\pi^2T^2\over 3}\Big)$. 

In Eq.(\ref{eqno(3)}), fermion propagator has the following spectral representation with the notation 
$\bf{k}=(\bf{p}-\bf{q})$ \cite{BELLAC},

\begin{equation}
S_f(i\omega_n,{\bf k})=\int_{-\infty}^\infty{{\rm d}k_0\over
2\pi}\,{( {K\llap{/\kern3pt}}+m)\rho_f(K)\over
k_0-i\omega_n-\mu} \ .
\label{eqno(4)}
\end{equation}
Hence, for $\rho_f(K)$ we use the free spectral density given by,
\begin{equation}
\rho_f(K)={\pi \over E_k}[\delta(k_0-E_k)-\delta(k_0+E_k)] \ .
\label{eqno(5)}
\end{equation}

One can take the imaginary part of the Eq.(\ref{eqno(3)}) to calculate the scattering rate with the help of Eq.(\ref{eqno(2)}).
For the calculation of the drag coefficient, one then inserts the energy exchange $\omega$ in the expression of 
$\Gamma$ and calculate $dE/dx$ from Eq.(\ref{eqno(1)}) to obtain,

\begin{eqnarray}
 -{dE\over dx}&=&{\pi e^{2}\over Ev_i} \int{{\rm d}^3q\over
(2\pi)^3}\int_{-\infty}^\infty{{\rm d}k_0\over 2\pi}\rho_f(k_0)
\int_{-\infty}^\infty{{\rm d}q_0\over 2\pi}q_{0} \nonumber\\
&\times &
(1+ n(q_0)- {\bar n}(k_0))  \delta(E-k_0-q_0)\nonumber \\
&\times &[p_0k_0+{\bf p\cdot
k}+m^2] 
  \rho_{l}(q_0,q) \nonumber \\
&+&2[p_0k_0-({\bf p\cdot \hat q})({\bf k\cdot\hat
q})-m^2]\rho_{t}(q_0,q). 
\label{eqno(8)}
\end{eqnarray}
It is to be mentioned here that the scattering process involves space like photons. Hence,
 here only the cut of the spectral function contributes. In the above equation $n$ and 
${\bar n}$ are the Bose-Einstein and the Fermi-Dirac
distribution functions:
\begin{equation}
n(q_0)={1\over {\rm e}^{\beta q_0}-1} \ , \qquad
{\bar n} (k_0)={1\over {\rm e}^{\beta (k_0-\mu)}+1} \ .
\label{eqno(10)}
\end{equation}

From now onwards in this section, we exclusively focus on the 
ultradegenerate plasma. The finite temperature corrections which
might be important for dense and warm plasma will be incorporated in
the next section. For $T=0$, $\mu\neq0$ limit, $(1+n(q_0))=\Theta(q_0)$ and
${\bar  n}(k_0)=\Theta(\mu-E+q_0)$, where $\Theta$ represent the step 
function. These functions, as we shall see, restrict
the phase space of the $q_0$ integration severely. The zero temperature spectral functions $\rho_{l,t}$ now involve Debye mass
 $m_D^2=e^2v_f\mu^2/\pi^2$.

Note that, the delta function in Eq.(\ref{eqno(8)}) sets $q_0=qvcos\theta$ and the theta functions 
impose further restrictions on $q_0$. We consider quasiparticles, with velocity close to Fermi velocity, which undergo collisions with the particles 
near the Fermi surface. Hence, we can make an approximation here as 
$v\approx v_f$. 

Now, consider the case of hard photon exchange where
the medium effects on the photon propagator can be ignored. In this
case using the bare propagator we get,

\begin{eqnarray}
 \Big(-{dE\over dx}\Big)
&\simeq&  {e^2 m_D^2  \over 8 \pi v_f }    
\int{  dq } \int_{0}^{E-\mu} d q_0 \left\{ 
 \frac{q_0^{2}}{ v_f^2 q^4  }
 +  \frac{v_f^2q_0^{2}} {  2 \, q^4  }
 \right\}  \ \nonumber\\
&  \simeq & {e^2 (E-\mu)^3
  m_D^{2}\over 24\pi v_f}({1\over v_f^2}+{v_f^2\over 2})\int{{dq}\over q^4}.
\label{eqno(19)}
\end{eqnarray}

This actually is the leading hard contribution that comes from the diagram, when, the blob of Fig.(\ref{fig1}) is replaced with 
one fermion loop. Evidently, the above integral is infrared 
divergent and unlike the finite temperature here higher powers
of $q$ appear in the denominator. We shall remark on this later once
we have expressions both for $\eta$ and ${\cal B}$.

To deal with this infrared divergence in the 
soft domain, one uses HDL corrected
photon propagator \cite{MANUEL2, MANUEL3} given by Eq.(\ref{eqno(7)}), with Debye mass $m_D^2=e^2v_f\mu^2/\pi^2$ as 
mentioned earlier, to obtain, 

\begin{eqnarray}
\Big(-{dE\over dx}\Big)\Big|_{\rm soft}(E) 
& \simeq & {e^2\over 2 v_f} \int \frac{{\rm d}^3q}{(2\pi)^3} qv_f\cos\theta  
( \Theta(q_0) \, -\Theta(\mu-E+q_0) ) \,\nonumber \\ 
&  \times &\Theta(q^*-q) \{ \rho_l(q_0, q)
+v_f^2(1-\cos^2\theta)\rho_t(q_0,q) \}\nonumber\\
& \simeq & {e^2\over 8 \pi ^2v_f} \int_D {\rm d}q q \int {\rm d}q_0 q_0  
\{ \rho_l(q_0, q)
+(v_f^2-\frac{q^2_0}{q^2})\rho_t(q_0,q) \}
 \ .
\label{eqno(11)}
\end{eqnarray}

The integration domain($D$) above is limited by the $\Theta$ functions, 
\begin{eqnarray}
D: &0&\leq q_0\leq E-\mu;\nonumber\\
  &q_0&\leq q\leq q^*  
\label{eqno(14)}.
\end{eqnarray}

With these we get, 
\begin{eqnarray}
 \Big(-{dE\over dx}\Big)\Big|_{\rm soft}(E)  &\simeq& {e^2 m_D^2  \over 4 \pi v_f}   
 \int_{D}  dq_0  dq  \nonumber\\
  &\times& \Big\{
 {q_0^{2}\over v_f^{2}\{2 \left[q^2 + m_D^2 Q_l({q_0\over q}) \right]^2  +
 {m_D^4 \pi^2 q_0^2 \over  2q^2}\}}  
\nonumber \\
 &+&{v_f^2 q_0^{2}\over\left[ 2q^2 + m_D^2 v_f^2Q_t({q_0\over q})  \right]^2  +
 {m_D^4 v_f^4\pi^2 q_0^2 \over 4q^2 }}\Big\},
\label{eqno(12)}
\end{eqnarray}
 where,  
\begin{equation}
Q_l(x)  =  1 -  \frac{x }{2}
\ln{\frac{1+x} {1- x} } \ ,
\qquad 
Q_t (x)  =  - Q_l (x) + {1 \over 1 - x^2  } \ .
\label{eqno(13)}
\end{equation}
We are mainly interested in the energy loss of a quasiparticle which is present close to the Fermi 
surface, hence, $(E-\mu)\ll m_D$ is the physically interesting region where the
quasiparticle concept is meaningful.
 The denominator of the Eq.(\ref{eqno(12)}) can now be expanded in powers of $q_0$. We replace $s^* =( q^*/ m_D)^2$ and compute separately the
electric ($l$) and the magnetic part ($t$),  
\begin{eqnarray}
\Big(-{dE\over dx}\Big)\Big|_{soft }^{l} &  \simeq & {e^2 (E-\mu)^3  \over 48 \pi m_Dv_f^3} 
\int_{0}^{s^*} {  ds  \over \sqrt{s} (s+1)^2 }     \ ,
\\
\Big(-{dE\over dx}\Big)\Big|_{soft }^{t} &  \simeq & {e^2 m_D^2 v_f^2 \over 4  \pi v_f} 
\int_{D}   dq_0 dq {q_{0}^{2}\over 4\,q^4+{\pi^2m_D^4 v_f^4q_{0}^2\over 4q^{2}}  }
\label{eqno(16)}. 
\ 
\end{eqnarray}
After explicit calculation, the electric and magnetic contributions to the expression of energy loss take the following form,
\begin{eqnarray}
\Big(-{dE\over dx}\Big)\Big|_{soft }^{l} 
&  \simeq & {e^2 (E-\mu)^3  \over 96 v_f^3 m_D}
-{e^2 (E-\mu)^3 m_D^{2} \over 72 \pi v_f^3 q^{\ast 3}}, 
\label{eqno(17)}
\end{eqnarray}
\begin{eqnarray}
 \Big(-{dE\over dx}\Big)\Big|_{soft }^{t} 
&  \simeq & {e^2 (E-\mu)^2 \over 48 \pi v_f }
-{e^2 (E-\mu)^3 v_fm_D^{2}\over 144\pi q^{\ast 3}}.
\label{eqno(18)}
\end{eqnarray}

It is worthwhile to note here that the leading order terms in the last
two equations are finite and independent of the cut-off parameter. 
Here the $q^*$ dependent term appear only at  $O(e^4)$. 
Therefore, we write,

\begin{eqnarray}
\Big(-{dE\over dx}\Big)\Big|_{soft }^{l} 
&  \simeq & {e^2 (E-\mu)^3  \over 96 v_f^3 m_D}+ O(e^4),
\label{eqno(17a)}
\end{eqnarray}
\begin{eqnarray}
 \Big(-{dE\over dx}\Big)\Big|_{soft }^{t} 
&  \simeq & {e^2 (E-\mu)^2 \over 48 \pi v_f } + O(e^4).
\label{eqno(18a)}
\end{eqnarray}

So far we have discussed about the soft part and have seen that in
 the limit $q^*\rightarrow \infty$ the cutoff
parameter dependent term trivially vanishes. Similarly, if we recall
the expression for the hard part {\em i.e.} Eq.(\ref{eqno(19)}),
after performing the integration  
in the limit $[q^{\ast}, \mu ]$ we get, 

\begin{eqnarray}
 \Big(-{dE\over dx}\Big)\Big|_{hard }
&\simeq& {e^2 (E-\mu)^3
  m_D^{2}\over 72\pi v_f}({1\over v_f^2}+{v_f^2\over 2})\Big [{1\over q^{\ast 3}}-{1\over \mu^{3}}\Big]\nonumber\\
&\simeq & O(e^4).
\label{eqno(19a)}
\end{eqnarray}
Clearly it fails to contribute at the leading order where the
entire contribution comes from the soft sector.
This is a distinctive feature of degenerate plasma not encountered at finite 
temperature ($\mu=0$). There
both the hard and the soft part contribute to the leading order in $e^2$ and
the divergence is only logarithmic. 
To deal with such divergences in hot plasma one invokes Braaten and 
Yuan's prescription \cite{BY} where an intermediate
cutoff is introduced to separate the hard and the soft domains. It is seen
that such an intermediate cutoff parameter disappears from the final
expressions when both the contributions are added.  
At zero temperature, a similar approach was adopted for the calculation of 
fermion damping rate \cite{CRISTINA} where it was shown 
that such cancellation takes place also in degenerate plasma. 
It is obvious from Eqs. ({\ref{eqno(17)}, \ref{eqno(18)}, \ref{eqno(19a)}}) that same 
thing happens for $\eta$ also.  

From Eq.(\ref{eqno(19a)}) it is clear that the result obtained from the hard region is 
suppressed with respect to the soft one (Eq.(\ref{eqno(17)}) and (\ref{eqno(18)})). 
Hence, the whole contribution to leading order 
comes from the soft sector alone. The final expression for drag coefficient at zero temperature becomes,
\begin{eqnarray}
 \eta&\simeq&{e^2 (E-\mu)^3  \over 96 m_D v_f^4 E}+{e^2 (E-\mu)^2 \over 48 \pi v_f^2 E}
+O(e^4).
\label{eqno(20a)}
\end{eqnarray}

The first term above corresponds 
to the electric photon and the latter to the magnetic one {\em i.e.} $l$ or $t$ mode behaves differently. The dominant contribution to $\eta$ comes from the magnetic sector in the ultrarelativistic case $v_f\rightarrow 1$ and the electric 
sector when $v_f\ll 1$. Results for the light fermion can be obtained 
from Eq.(\ref{eqno(20a)}) with the substitution of $v_f\rightarrow1$. 

%**************************************************************************************************
%**************************************************************************************************

\subsection{Diffusion coefficient}

%*************************************************************************************************
%*************************************************************************************************

Apart from $\eta$, the quantity which could be of importance in the study of heavy fermion propagating in the plasma is the 
momentum diffusion coefficient ($B_{ij}$) \cite{BENJAMIN, GDM, ABHEEP, BERAUDO}. In fact, we know for Coulomb plasma $\eta$ and the longitudinal momentum 
diffusion coefficient (${\cal B}$) are related {\em via} ER. Momentum diffusion coefficient ${B_{ij}}$ can be defined as follows \cite{BENJAMIN,ABHEEP},
\begin{eqnarray}
 { B_{ij}}=\int d\Gamma q_i q_j.
\end{eqnarray}
Decomposing $B_{ij}$ into longitudinal ($B_l $) and transverse components ($B_t$) we get the following expression,
\begin{eqnarray}
 B_{ij}=B_t (\delta_{ij}-\frac{p_i p_j}{p^2}) + B_l \frac{p_ip_j}{p^2}.
\end{eqnarray}
These coefficients $B_{l,t}$ are the longitudinal, transverse squared momentum acquired by the particle through collision 
with the plasma. Using the above definition, like the energy loss (Eq.(\ref{eqno(8)})), longitudinal momentum diffusion coefficient ($ B_l={\cal B}$, suppressing the index $l$) can be written as follows,
\begin{eqnarray}
 {\cal B}&=&{\pi e^{2}\over E} \int{{\rm d}^3q\over
(2\pi)^3}\int_{-\infty}^\infty{{\rm d}k_0\over 2\pi}\rho_f(k_0)
\int_{-\infty}^\infty{{\rm d}q_0\over 2\pi}q_{||}^2 \nonumber\\
&\times &
(1+ n(q_0)- {\bar n}(k_0))  \delta(E-k_0-q_0)\nonumber \\
&\times &[p_0k_0+{\bf p\cdot
k}+m^2] 
  \rho_l(q_0,q) \nonumber \\
&+&2[p_0k_0-({\bf p\cdot \hat q})({\bf k\cdot\hat
q})-m^2]\rho_t(q_0,q). 
\label{eqno(27)}
\end{eqnarray}
Here, $q_{||}=q {\rm cos} \theta $. For the exchange of hard photons using the bare propagator we obtain: 
\begin{eqnarray}
 {\cal B}&  \simeq & {e^2 m_D^2  \over 8 \pi v_f^2 } ({1\over v_f^2}+{v_f^2\over 2})
\int dq \int_{0}^{E-\mu}   dq_{0}  {q_{0}^{3}\over q^{4}},\nonumber\\
 &\simeq& {e^2 m_D^2(E-\mu)^{4}  \over 32 \pi v_f^2 }({1\over v_f^2}+{v_f^2\over 2}) \int \frac{dq}{q^{4}}.
\label{eqno(31)}
\end{eqnarray}

Comparing Eqs.(\ref{eqno(19)}) with (\ref{eqno(31)}) it is seen that like $\eta$, ${\cal B}$ is also infrared divergent involving fourth power
of $q$ in the denominator. At finite $T$ case, both the quantities are proportional
to $dq/q$ at the
leading order \cite{ABHEEP}. To understand the origin of this difference,
we focus on the $q_0$ integration. It is shown in the appendix that in
medium at finite $T$, there involves quadratic power of $q_0$ in both cases with
the limits $-vq$ to $+vq$ giving rise to a term proportional to
$q^3$ in the numerator. This cancels with some of the powers of $q$
coming from the propagator. Whereas in cold matter, from Eqs.(\ref{eqno(19)}), (\ref{eqno(31)}) we
find that the same integrations appear with $q_0^2$ and $q_0^3$
in the numerator while the limits are independent of $q$ forbidding
the cancellation with $q$'s coming from the propagator as before.
We note here that 
the drag and diffusion coefficients are related through 
${\cal B}= {3E(E-\mu)\over 4}\eta$ when we deal with the bare propagator. We shall see in the next paragraph that same powers of $q_0$ appear in the numerator when one takes the plasma 
effects into account but such common scaling behavior is lost.

The infrared  divergence of Eq.(\ref{eqno(31)}) can be removed by using the dressed photon propagator \cite{MANUEL2, MANUEL3} and providing the upper
 cut off as in the case of $\eta$. With the HDL corrected propagator one gets,
\begin{eqnarray}
{\cal B}\Big|_{\rm soft}(E) 
& \simeq & {e^2\over 2v_f^2 }  \int \frac{{\rm d}^3q}{(2\pi)^3}
q_0^2  
( \Theta(q_0) \, -\Theta(\mu-E+q_0) ) \,\nonumber \\ 
&  \times &\Theta(q^*-q) \{ \rho_l(q_0, q)
+v_f^2(1-\cos^2\theta)\rho_t(q_0,q) \}\nonumber\\
& \simeq & {e^2\over 8 \pi ^2v_f^2} \int_D {\rm d}q q \int {\rm d}q_0 q_0^2  
\{ \rho_l(q_0, q)
+(v_f^2-\frac{q^2_0}{q^2})\rho_t(q_0,q) \}
 \ ,
\label{eqno(11a)}
\end{eqnarray}
here the integration domain ($D$) is same as before.
So,
\begin{eqnarray}
{\cal B}|_{\rm soft}(E)  &\simeq& {e^2 m_D^2  \over 4 \pi  v_f^{4}}   
 \int_{D}  dq_0  dq  \nonumber\\
  &\times& \Big\{
 {q_0^{3}\over\{2 \left[q^2 + m_D^2 Q_l({q_0\over q}) \right]^2  +
 {m_D^4 \pi^2 q_0^2 \over 2q^2}\}}  
\nonumber \\
 &+&{v_f^4 q_0^{3}\over\left[ 2q^2 + m_D^2 v_f^2Q_t({{q_0\over q}})  \right]^2  +
 {m_D^4 v_f^4\pi^2 q_0^2 \over 4q^2 }}\Big\}.
\label{eqno(34)}
\end{eqnarray}
Since, we know from energy loss that dominant contribution to the expression comes from the soft region alone we write the 
expression for ${\cal B}$ as follows,
\begin{eqnarray}
{\cal B}&\simeq&{e^2 (E-\mu)^4  \over 128 m_{D} v_f^4}+{e^2 (E-\mu)^3 \over 72 \pi v_f^2}+O(e^4),
\label{eqno(36)}
\end{eqnarray}
which is finite. Now from Eqs.(\ref{eqno(20a)}) and (\ref{eqno(36)}) it can be seen that there is 
no common scaling factor between $\eta$ and ${\cal B}$. But as ER is formulated in the region where
 $v_f\ll 1$, in this nonrelativistic region exchange of the magnetic photons are suppressed in comparison with the electric one. Hence, considering only the electric part we get
 the same ER, ${\cal B}= {3E(E-\mu)\over 4}\eta$ as in the case of bare perturbation theory.

%**********************************************************************************************************
%**********************************************************************************************************

\section{Finite temperature correction}

%**********************************************************************************************************
%**********************************************************************************************************
The results of the previous section can easily be extended to the case of a hot and dense ($T\ll\mu$) plasma. This could be relevant for heavy ion collision to be performed
 at GSI where the chemical potential is expected to be much higher than the temperature. Now, while calculating the soft part we replace the zero temperature 
distribution functions with the finite temperature one in Eq.(\ref{eqno(11)}) and write,
\begin{eqnarray}
\Big(-{dE\over dx}\Big)\Big|_{\rm soft}(E) & \simeq & {e^2\over 2v_f} \int {{\rm d}^3q \over
(2\pi)^3}q v_fcos\theta ( 1+n(q_0)-{\bar n}(E-q_0) ) \,\nonumber \\ 
&  \times &\Theta(q^*-q) \{ \rho_{l}(q_0, q)
+v_f^2(1-\cos^2\theta)\rho_{t}(q_0,q) \}
 \ .
\label{eqno(22)}
 \end{eqnarray}

With small $T$ and large $\mu$, the above equation can be 
calculated according to Ref.\cite{FEYNMAN}. In this approach we can write any function 
$g(\varepsilon)$ along with the fermion distribution function as follows, 
\begin{eqnarray}
\int_0^\infty {g(\varepsilon)\over  {\rm e}^{\beta (\varepsilon-\mu)}+1}d\varepsilon =\int_0^\mu g(\varepsilon)d\varepsilon+
{\pi^2T^2\over 6}g'(\mu).
\label{eqno(23b)}
\end{eqnarray}

The contributions
coming from soft ($l$ and $t$) using Eqs.(\ref{eqno(22)}, \ref{eqno(23b)}, \ref{eqno(9)}) are found to be given by:
\begin{eqnarray}
 \Big(-{dE\over dx}\Big)\Big|_{soft }^{l} &  \simeq & {e^2 (E-\mu)^3  \over 96 v_f^3 m_{D}}
-{e^2 (E-\mu)^3 m_{D}^{2}\over 72\pi q^{*3} v_f^3} 
- {e^2 (E-\mu)T^2\pi^2 \over 96  v_f^3m_{D} }
+{e^2 (E-\mu)T^2m_{D}^2\pi\over 72 q^{*3}v_f^3}, 
\label{eqno(23a)}
\end{eqnarray}
\begin{eqnarray}
 \Big(-{dE\over dx}\Big)\Big|_{soft }^{t} &  \simeq & {e^2 (E-\mu)^2\over 48 \pi  v_f}
-{e^2 (E-\mu)^3 v_f m_{D}^{2}\over 144\pi q^{\ast 3}}
-{e^2 \pi T^2\over 72 v_f}
+{e^{2} (E-\mu) m_{D}^{2}v_fT^{2}\pi \over 144 q^{\ast 3}}.
\label{eqno(24)}
\end{eqnarray}
We see from the above two equations that the terms containing separation scale are subleading 
in comparison with the others, the same behavior obtained in the zero temperature case also.
The term with the bare propagator comes as,
\begin{eqnarray}
 \Big(-{dE\over dx}\Big)\Big|_{hard }& \simeq &{e^2 (E-\mu)^3
  m_{D}^{2}\over 72\pi v_f}({1\over v_f^2}+{v_f^2\over 2})\Big [{1\over q^{\ast 3}}-{1\over \mu^{3}}\Big]\nonumber\\
&-&{e^2 (E-\mu)m_{D}^2\pi T ^{2}\over 72  v_f}({1\over v_f^{2}}+{v_f^2\over 2})\Big[{1\over q^{\ast 3}}-{1\over \mu^{3}}\Big].
\label{eqno(25)}
\end{eqnarray}

The above expression is also suppressed in contrast to the soft one. Hence, with finite temperature correction, drag and diffusion coefficients become, 
\begin{eqnarray}
\eta&\simeq&{e^2 (E-\mu)^3  \over 96 m_{D} v_f^{4}E}+{e^2 (E-\mu)^2  \over 48 \pi v_f^2E }-{e^2 \pi^{2}T^{2}(E-\mu)  \over 96 m_{D} v_f^{4}E}-{e^2 \pi T^2\over 72 v_f^2 E}+O(e^4),
\label{eqno(26a)}
\end{eqnarray}
\begin{eqnarray}
{\cal B}&\simeq&{e^2 (E-\mu)^4  \over 128 m_{D} v_f^4}+{e^2 (E-\mu)^3 \over 72 \pi v_f^2}-{e^2 (E-\mu)^2\pi^2 T ^{2}\over 64 m_{D}v_f^4}-{e^2 \pi T^2(E-\mu)\over 48 v_f^2}+O(e^4).
\label{eqno(38)}
\end{eqnarray}
One notes here with the thermal correction the ER cannot be established even for the electric 
sector alone.

\section{Summary}

In this work we calculate the energy loss and momentum diffusion
of heavy fermion in dense and warm QED matter and highlight some
of the differences that exist between the hot ($\mu=0$) and the cold ($T=0$) plasma. Unlike finite temperature, where one encounters logarithmic
divergences in calculating $\eta$ or ${\cal B}$, here we come across 
non-logarithmic divergences. Furthermore,
we see that at the leading order in coupling, the entire contribution  
comes from the soft sector and this is finite {\em i.e.} the physics 
here is dominated by the excitations near the Fermi surface. The exchange
of hard photons on the other hand contribute only at $O(e^4)$. It is
to be noted that in a thermal medium with vanishing chemical potential
both the soft and hard photons or gluons for QCD matter contribute
at the same order. Moreover, for ultrarelativistic particles both $\eta$ and ${\cal B}$ receive
dominant contributions from the magnetic sector while the
electric parts are  found to be subleading in $(E-\mu)$. 
This is consistent with fermion damping rate calculation 
\cite{CRISTINA, MANUEL2, VO} in degenerate plasma. Quantitatively, we find
that for the 
transverse or magnetic interaction $\eta$ is proportional to $(E-\mu)^2$ 
while for the electric interaction, it goes as $(E-\mu)^3 $. 
Similar differences for ${\cal B}$ is also seen where one more 
extra power of $(E-\mu)$ involved in each case. The other important finding
of the
present investigation is the ER for the drag and diffusion
coefficient. In hot plasma, it is known that 
${\cal B}=2TE\eta$ \cite{GDM, ABHEEP, BERAUDO}. At zero temperature, we find, 
${\cal B}={3E(E-\mu)\over 4}\eta$ by considering only the bare propagator
{\em i.e.} when we do not take the plasma effects into account. However,
we see that this common scale behavior is lost for
soft photon exchange where the plasma effects are
included and both the magnetic and electric contributions are retained.
However, by retaining only the electric contribution for the cold
plasma, one arrives at the same relations as obtained by using the
bare propagator. For $T\ll\mu$ again we see that, with plasma
effects incorporated, $\eta$ and ${\cal B}$ fail to show such
common scale behavior even when the magnetic interaction is ignored.

As a last remark, we note that here the entire calculation has been
done for QED plasma. This can easily be extended to QCD matter with
appropriate modifications like inclusion of diagrams involving three
gluon interaction and proper vertex factors coming from the QCD color algebra.
Such studies are in progress and shall be reported in future.

% *************************************************************************************************************************
% *************************************************************************************************************************

\section{Acknowledgment}

% *************************************************************************************************************************
% *************************************************************************************************************************

We would like to thank the referees for their useful comments and
S. Mrowczynski for fruitful communications. We also acknowledge 
helpful discussions with K. Pal and P. Roy.
\section {Appendix}
To understand the difference of the results between the cold and hot plasma, 
we first recall the expression  for the 
drag coefficient ($\eta$):
\begin{eqnarray}
\eta&=& {1\over E v_i}\Big(-{dE\over dx}\Big).\nonumber\
\end{eqnarray}
The above relation with Eqs.(\ref{eqno(4)}), (\ref{eqno(5)}) and (\ref{eqno(8)}) can be further simplified to yield,
\begin{eqnarray}
\eta &  \simeq & {e^2\over 8 \pi ^2v^2 E} \int {\rm d}q q \int_{-vq}^{vq}{\rm d}q_0 q_0 
(1+ n(q_0)- {\bar n}(k_0)) \Theta(q^*-q) 
\{ \rho_l(q_0, q)
+(v^2-\frac{q^2_0}{q^2})\rho_t(q_0,q) \}.
\label{eqno(41a)}
\end{eqnarray}

The corresponding expression for the diffusion coefficient from Eq.(\ref{eqno(27)}) is,
\begin{eqnarray}
 {\cal B}&\simeq & {e^2\over 8 \pi ^2v^2 } \int{\rm d}q q \int  _{-vq}^{vq}{\rm d}q_0 q_0^2 
(1+ n(q_0)- {\bar n}(k_0)) \Theta(q^*-q) 
\{ \rho_l(q_0, q)
+(v^2-\frac{q^2_0}{q^2})\rho_t(q_0,q) \}.
\label{eqno(43a)}
\end{eqnarray}

In the high temperature limit, $(1+ n(q_0)- {\bar n}(E-q_0))\simeq 
{T\over q_0}+{1\over 2}$. It is to be noted that the above integration limits are
symmetric in $q_0$. Hence, for the drag,  the factor of ${1\over 2}$ 
and for the diffusion ${T\over q_0}$ contribute. Inserting  these in 
Eqs.(\ref{eqno(41a)}) and (\ref{eqno(43a)}) we get,
\begin{eqnarray}
\eta &  \simeq & {e^2\over 16 \pi ^2v^2 E} \int {\rm d}q q \int_{-vq}^{vq}{\rm d}q_0 q_0 \Theta(q^*-q) 
\{ \rho_l(q_0, q)
+(v^2-\frac{q^2_0}{q^2})\rho_t(q_0,q) \},
\label{eqno(42a)}
\end{eqnarray}
and
\begin{eqnarray}
 {\cal B}&\simeq & {e^2 T\over 8 \pi ^2v^2 } \int{\rm d}q q \int  _{-vq}^{vq}{\rm d}q_0 q_0\Theta(q^*-q) 
\{ \rho_l(q_0, q)
+(v^2-\frac{q^2_0}{q^2})\rho_t(q_0,q) \}.
\label{eqno(44a)}
\end{eqnarray}
In case of bare interaction, one can show that both $\eta$ and ${\cal B}$
 is proportional to $d{\rm q}/ q$ \cite{ABHEEP} and even without performing the integration ${\cal B}=2TE\eta$. If we compare the 
Eqs.(\ref{eqno(42a)}) and (\ref{eqno(44a)}) with Eqs.(\ref{eqno(11)}) and (\ref{eqno(11a)}) we find that 
the $q_0$ integration for cold matter is not symmetric, and the limits are independent of $q$. Here, lies the difference of cold and hot plasma.

%%%%%%%%%%%%%%%%%%%%%%%%%%%%%%%%%%%%%%%%%%%%%%%%%%%%%%%%%%%%%%%%%%%%%%%%%%%%%%%
\end{document}